# EDOCR: ENERGY DENSITY ON-DEMAND CLUSTER ROUTING IN WIRELESS SENSOR NETWORKS

B M Thippeswamy[1], Reshma S [2], Shaila K [2], Venugopal K R [2], S S Iyengar[3], L M Patnaik[4]

[1] Department of Computer Science and Engineering, Jawaharlal Nehru Technological University, Anantpur.
[2] Department of Computer Science and Engineering,
University Visvesvaraya College of Engineering, Bangalore University, Bangalore.
[3]Ryder Professor, Florida International University, U S A.
[4]Honorary Professor, Indian Institute of Science, Bangalore, India.

## ABSTRACT

*Energy management is one of the critical parameters in Wireless Sensor Networks. In this paper we attempt for a solution to balance the energy usage for maximizing the network lifetime, increase the packet delivery ratio and throughput. Our proposed algorithm is based on Energy Density of the clusters in Wireless Sensor Networks. The cluster head is selected using two step method and on-demand routing approach to calculate the balanced energy shortest path from source to sink. This unique approach maintains the balanced energy utilization among all nodes by selecting the different cluster heads dynamically. Our simulation results have compared with one of the plain routing scheme (EBRP) and cluster based routing (TSCHS), which shows the significant improvements in minimizing the delay and energy utilization and maximizing the network lifetime and throughput with respect to these works.*

## 1. INTRODUCTION

Wireless Sensor Networks(WSNs) are used in numerous applications like Traffic management, Battle field surveillance, Environmental monitoring, Health care systems, Underwater applications and etc., [1][2][3].

Energy utilization is one of the significant parameter for battery powered wireless sensor networks. It is essential to reduce energy consumption in all the sensor nodes to increase the network lifetime [4]. In WSNs, the nodes surrounding the sink have tendency to drain their energy soon compared to the nodes away from the sink and such irregular energy drain will decrease the network lifetime [5]. Unbalanced energy utilization can cause network partition even though many of the nodes may have maximum residual energy which are away from the sink [6]. Thus, it is necessary that every node should consume energy evenly in order to increase the lifetime of the network.

Energy efficiency and balanced energy utilization are two different aspects. Shortest path routing uses energy efficiently but may not result in balanced energy utilization. Topology, Applications and Routing protocols are main causes for unbalanced energy utilization. However, eventually a number of solutions are proposed by many of the routing protocols such as : Optimal Deployment of sensor nodes relative to applications [7][8], Organization of dynamic Topology of nodes based on transmission power requirements [9][10], the deployment of Mobile sinks or Relay nodes [11][12] and efficient data aggregation techniques to manage uniform energy utilization across all nodes in the network [13][14].

## 1.1. Motivation

Wireless Sensor Networks are battery powered and hence the scarce energy is an important and has to be utilized effectively and efficiently. The goal of the WSNs is to increase its lifetime and maximize its throughput. In earlier works, cluster based routing protocols have used different parameters such as residual energy, distance from cluster head to the base station, etc to elect the cluster head. The cluster head drains out energy faster than all other nodes in the cluster. And more over density of clusters may vary in the network. Hence it is quiet important to consider residual energy of all the nodes in the cluster adoptively. In addition, the density of nodes in a cluster plays the critical role in overall average energy of the cluster, a parameter that needs to be computed for the disconnection of the cluster resulting in the partition of the network. There must be a mechanism which enables uniform utilization of energy across all nodes in the cluster. And as well as use those nodes along with a group of neighbouring nodes with higher residual energy as cluster head to increase the lifetime and minimize the throughput in the network.

## 1.2. Contribution

We have designed a cluster based routing protocol to utilize energy efficiently and effectively among all the nodes in a given cluster. The cluster head is chosen based on two parameters (i) the residual energy (ii) energy density of a given node. The energy density of each node is based on the ratio of sum of the residual energy of the neighbour nodes(including itself) and the distance from the local cluster head and coverage area. On-demand routing is employed through the cluster head to obtain the balanced energy shortest path. This approach consumes lower energy and utilizes energy uniformly across all nodes in the entire network resulting in maximizing lifetime and throughput.

## 1.3. Organization:

This paper is organized as follows. Section II presents a brief literature of related works. Section III discusses the background work, while section IV defines the problem and objectives. Section V describes system and mathematical model and proposed algorithm. Section VI presents the performance evaluation. The concluding remarks are summarized in section VII.

## 2. LITERATURE SURVEY

This section presents a brief summary of related works. Xu et al., [15] explored the idea of Geographical Adaptive Fidelity (GAF). This algorithm minimizes the energy consumption in ad-hoc wireless sensor network. Energy conservation is done by identifying the important nodes with respect to routing perspective and turning off the unnecessary nodes. GAF is adaptive fidelity technique which extends the lifetime and self configuring systems. This algorithm does not address mobility model. It has to be tested under heavier traffic load.

Rodopher et al., [16] describes a distributed position based network protocol that effectively minimizes the energy consumption in mobile wireless networks and supports peer-to-peer communications. For the unlimited random deployment of nodes, each node guarantees the strong connectivity of the network and achieves global minimum energy distribution for stationary networks. The localized nature of this protocol is self-reconfiguring and gives minimum energy solution for mobile networks. This protocol has not considered all the potential nodes in a network while finding global minimum path.

Younis et al., [17] proposed a novel distributed clustering approach for ad-hoc sensor networks. This approach has assumed that multiple power level nodes are available in the network and it does not consider the infrastructure and node capability. Here, the selection of the cluster head is based on residual energy and node proximity to its neighbour. It achieves uniform cluster head distribution across the network and effectively prolonging the network lifetime. It also supports the scalable data aggregation. This protocol is designed for only two level hierarchy.

Singh et al., [18] focus on design of an energy balanced and energy optimal algorithm for sorting in a single-hop sensor network. The energy optimality is achieved by maintaining the balanced energy dissipation among all the nodes. Energy optimality and energy balancing is demonstrated for single-hop, single-channel network of randomly distributed sensors. The process of sorting is performed for specific amount of time and energy with no sensor being aware more than stipulated time steps. This algorithm can be applied successfully to low cost paging channels, but it increases the overall execution time for other types of channels. It also requires more accurate time synchronization between sensors.

Rahul et al., [19] address the effectiveness of using lowest energy path. The frequent usage about such paths may not be optimal for long network lifetime and connectivity point of view. This scheme proposes energy aware routing that uses sub optimal paths to achieve significant gains. The nodes burn the energy in more balanced way through out the network and results in more graceful degradation of service with time. The probabilistic forwarding followed in this protocol may not always result in the better routing path.

Baek et al., [20] investigated the use of proactive multi-path to optimize trade-off between the energy cost of spreading traffic and improved spatial balance of energy burdens using stochastic geometric and queuing models. It is compared with shortest path routing scheme with high initial energy. This work involves the considerable overheads related to setting up multi-path routes. Nurhayati et al., [21] presented a Cluster Based Energy efficient location Routing Protocol (CLERP) adopting hierarchical structure method, multi-hop and location-based node in the network. This routing is based on clusters in which cluster head aggregates the data and sends it to the base station. Such an approach reduces the energy consumption in a large scale network for long distance wireless communication. The Cluster head selection criteria is based on only the residual energy and distance from the base station that limits the level of energy balancing in the network.

Bager et al., [22] proposed an idea of cluster based routing protocol to extend the network lifetime: the cluster head selection is based on the residual energy. The aggregated data is sent to the base station by constructing the spanning tree. This process can handle the heterogeneous energy capacities and prolong the network lifetime. The spanning tree approach for data transmission is concentrated only on optimization of energy consumption by the cluster heads but not energy balancing.

Ashok et al., [23] proposed a location-based protocol for WSN which supports an energy-efficient clustering, cluster head selection/rotation and data routing methods to extend network lifetime. Clustering ensures the balanced size cluster formation within the boundary of sensing field with minimum number of transmit and receive operations. Then Cluster head rotation ensures balanced energy dissipation of the node in spite of the non-uniform energy requirement of cluster head sensor nodes. In this scheme, the cluster head rotation will be more effective by considering the constraints like distance and energy density rather than residual energy.

Wang et al., [24] presented distance based clustering routing protocol, (LEACH-Selective cluster). In this approach every node checks for the cluster head which is closest to the centre point to itself and sink, that is, every node has got the freedom to select the cluster head of different clusters. Thus, the overall network energy consumption is reduced considerably and there is significant improvement in network lifetime. The major limitation of this approach is that all the nodes must have their location information in WSN.

Pan et al., [25] describes a cluster based routing protocol that adds a tiny slot in a round frame, which facilitate the exchange of residual energy messages between base station, cluster heads and nodes. The cluster head gathers the energy status of all the nodes that belongs to it and sends it to the base station. In turn the base station computes and broadcast the overall residual energy to the cluster heads. Then, all the nodes in the network receive this information through their respective cluster heads and they get ready for the next round. This process makes all the nodes including cluster heads in a balanced manner to take collective decision at every round. This approach can get affected by critical overheads involved in residual energy information exchanges for networks.

Huabiao et al., [26] proposed an improvement in LEACH where the multi-hops path constructs by the gateways are used for the packet transmission. The cluster heads are restricted to gathering and aggregating the data from cluster members. The energy balance of the cluster heads is not depleting, as the transmission process is completely detached from them. The complexity involved in multi-hops path selection can increase the time overhead and it may degrade the overall performance.

Bochang et al., [27] developed an Energy-efficient Cluster based Data-Gathering Protocol (ECDGP). It involves a method known as belief degree for cluster head selection. The active nodes are selected according to the demand of network coverage. This method of controlling the number of active nodes in a cluster reduces the energy consumption and enhances the network lifetime. The belief degree based cluster head solution is found to be more application specific.
Xiaorong et al., [28] used static cluster formation based on node locations, communication efficiency and network connectivity. The cluster head selection is based on optimal scheduling that enhances the lifetime of the cluster head for longer time. The cluster heads are responsible for periodical collection of data aggregation and forwarding using minimal energy routing. This approach effectively utilizes energy in a balanced manner and extends the network lifetime. Here cluster head is compelled to follow the fixed transmission rate.

Dang et al., [29] investigated the distributed clustering scheme for delay tolerant mobile networks. The mobile nodes are distributively grouped based on similar mobility pattern into a cluster and can interchangeably be used for load balancing to improve the network performance. The nodal contact probabilities are improved using weighted moving average scheme. Finally, the gateways carry out the routing by exchanging the network information. This protocol improved the delivery ratio, minimized the overhead and end-to-end delay.

Sundara et al., [30] presented a survey of state-of-art routing techniques in WSNs. Stefanos et al., [31] describes the concept of Equalized Cluster Head Election Routing Protocol (ECHERP). This protocol models the network as linear system using Gaussian Elimination algorithm to choose the cluster head. The selection of cluster head is based on the node that can minimize total energy consumption in a cluster rather than high residual energy. This approach includes the multi-hop routing to transmit the data from source to base station. QoS Metrics and network time constraints are drawbacks of this work.

Rui et al., [32] proposed cluster head selection based on the state information of the neighbour nodes in a cluster instead of the phenomenon of blind node concepts. The cluster head selection is based on Particle Swarm Optimization (PSO). Here, overhead occurs due to the cluster reconstruction and change over from auxiliary cluster head to actual cluster head.

Jamping et al., [33] discussed the time based cluster head selection algorithm to enhance the performance of LEACH known as TB-LEACH. The cluster head selection is based on the random time interval based on random number assigned to each node. This concept adopted the balanced cluster distribution and the numbers of cluster formations are minimized compared to LEACH. The TB-LEACH effectively minimized the number of clusters formation and simplified the process of cluster head selection.

Anitha et al., [34] designed an Energy Efficient Cluster Head Selection (EECHS) protocol in mobile wireless sensor network. This protocol selects the cluster head nodes based on the residual energy, lowest mobility factor and density of the node. The gateways are used to transfer the data from nodes to the base station. This protocol minimizes the energy consumption and maximizes network lifetime and throughput. This approach involves the delay overheads as part of the data transmission due to the intermediate gateway nodes between the node and base station.

Zhong et al., [35] proposed a routing protocol called Two-Step Cluster Head Selection (TSCHS) to avoid the cluster head number variability problem of LEACH protocol. Here the cluster head is selected in two different stages. In the first stage, initially, the temporary cluster heads are selected and the numbers of cluster heads are more than optimal value. In the second stage, the optimal number of cluster heads are chosen based on the residual energy and the distance to the base station. Finally the numbers of temporary cluster heads are replaced with optimal value. Thus the network energy load is more balanced and prolongs the network lifetime. But for more effective energy balancing in the network, it is essential to consider one more prominent energy constraint like energy density along with the residual energy and the distance parameter.

Gao et al., [36] designed a Recluster-LEACH protocol based on the node density inside the cluster and cluster based data fusion mechanism. This protocol overcomes the limitations of single-hop LEACH protocol which consider the location information and the residual energy. The cluster head selection is based on multi-hop algorithm to enhance the energy efficiency and extend the network lifetime. There is a problem of greater energy depletion at the cluster heads near to the sink.

Jin-Su et al., [37] addressed the major challenges in WSN such as equal distribution of the clusters and the energy dissipation caused by the frequent information exchanged between the cluster heads and nodes. The cluster head is selected using the information received from upper level cluster head, distance among the nodes in the cluster, residual energy and density. This protocol also proposes the modified cluster head selection in which the conserved energy will be utilized in the steady state phase by minimizing unnecessary communications of unchanged nodes between selected cluster head and previous cluster head in the setup phase. It is necessary to minimize the delay in the setup phase that can manage the mobility among the nodes inside a network.

Sung-Ju et al., [38] performed the simulation and performance study of Dynamic Source Routing (DSR) and Associativity-Based Routing (ABR) for the multi-hop mobile wireless environment along with DSR. Dongkyun et al., [39] discussed the design of efficient Duplicate Address Detection (DAD) schemes for each node in a network. This protocol mainly focused on passive

DAD schemes over on-demand ad-hoc routing protocols with three main goals such as improving the accuracy of detecting address conflicts, improving the successful detection success ratio and reducing the time taken to detect these conflicts. This approach used the additional information like sequence number, location and neighbour knowledge which were not used in any previous approaches. It achieved better accuracy and shorter time to detect conflict addresses, but while resolving the address conflicts it is equally important to consider IP address allocation schemes.

David et al., [40] analyzed the use of on-demand behaviour of the routing protocols for the offered traffic load. The DSR has analyzed the latency and cost of the route discovery, and the effect of on-demand behaviour on routing cache consistency and also identified several mechanism that can be used to reduce the cost of route discovery. It is necessary to study the impact for variable number of nodes, the pattern of the node movement and type of communication pattern. Charles et al., [41] evaluated two dynamic routing protocols for ad-hoc networks such as DSR and AODV.

In our proposed protocol, we devised an efficient method of calculating the energy density of the cluster heads using two steps. In the first step, the local cluster head is selected based on the residual energy and the cost of the node. In the second step, the remaining cluster heads of different clusters are decided based on the energy density parameter of each node which is calculated as the ratio of sum of residual energy of neighbour nodes and itself to the distance from local cluster head and the coverage area. Finally, the shortest path from source to sink is decided based on On-Demand Routing.

## 3. BACKGROUND

Most of the existing energy efficient routing protocols forward the packets only through minimum energy path to minimize the energy consumption. But it causes an unbalanced distribution of residual energy in the network and results in early network partition.

In order to overcome this problem the Energy Balanced Routing Protocol (EBRP) [4] is designed by constructing a mixed virtual potential field in terms of depth, residual energy and energy density. The main goal of this technique is to force the packet to sent through the more energy densed area towards the sink, so that it can protect the nodes with relatively low residual energy. This protocol has considerably improved network lifetime, coverage ratio and throughput. But it can only find the route from each source node to a common sink and also requires to analyze the dynamics of time-varying potential field. Further it is necessary to check the looping problems to improve the integrated performance of the entire network.

Two-Step Cluster Head Selection (TSCHS) [35] is proposed to avoid the cluster head number variability problem of the LEACH protocol. Here, the cluster head is selected in two different stages. In the first stage, the temporary cluster heads are selected and the numbers of cluster heads are more than optimal value. In the second stage, the optimal number of cluster heads are chosen based on the residual energy and the distance to the base station. Finally, the temporary cluster heads are replaced with optimal value. Thus, the network energy load is more balanced and prolongs the network lifetime. In our work, we have identified the cluster head based on energy density, residual energy and distance. This method effectively balances the energy utilization, prolongs the network lifetime and maximizes the throughput.

## 4. PROBLEM DEFINITION

Given a wireless sensor network of $N$ nodes, we consider a two step cluster head selection process and routing through cluster head to utilize energy uniformly among all nodes. In the first step, the local cluster head $C_l$ is selected based on the residual energy $R_e$ and the cost of individual nodes in a network. In the second step, remaining cluster heads $CH_i$, where $i = 1, 2, 3. . . n$, of different clusters $C_i$, are selected based on the energy density $E_d$. After selecting the cluster heads, the shortest path request message is sent from event generated node to the sink through the neighbour cluster heads. In turn, the sink replies through the shortest path to the source, based on On-demand routing. Finally, the data is transferred from event generated node to the sink. The main objectives of our research work are:

(1) Uniform utilization of energy among all nodes to increase the lifetime of the network.
(2) Reduce packet drop ratio and increase throughput.

***Assumptions:***

(1) Initially energy levels of all the nodes are same.
(2) The sink possesses highest energy and is static.

## 5. SYSTEM AND MATHEMATICAL MODEL

The proposed model for EDOCR protocol is illustrated as shown in the Fig. 1. The system architecture is divided into five different phases. In the first phase, nodes are deployed randomly to create a network. The second phase involves cluster formation while the third phase generates the cluster head based on residual energy and energy density. On-Demand Shortest Path (ODSP) is used in the fourth phase based on the depth calculation. The fifth phase represents forwarding the packet from the source to the sink through the shortest path. The inputs to our system model are randomly deployed nodes. Here the cluster head selection phase and On-Demand Shortest Path (ODSP) are the two crucial phases. The following sub section describes the mathematical model and the different modules of the system architecture.

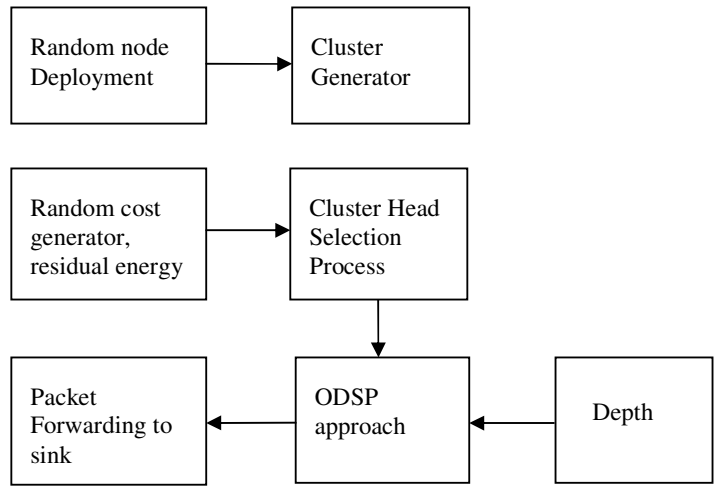

Figure 1 – Proposed model for EDOCR protocol

## 5.1. DEPTH

The depth field, $d$ is defined as the total number of cluster heads which come across every possible path between source and sink. Sink initializes its own depth, $d$ as $d = 0$, and broadcasts this value to its immediate neighbour cluster head. Then, the neighbour cluster head updates its depth as, $d = d + 1$. This process is repeated until all the cluster heads computes their depth $d$ to the sink.

## 5.2. RESIDUAL ENERGY

The residual energy, is defined as the energy retained in each node after the previous receiving and transmission process. The Residual energy $R_e$ can be computed as:

$$R_e = I_e - (E_t + E_r),$$ (1)

where, $I_e$ is the Initial Energy, $E_t$ and $E_r$ are the energy required for Transmission and Receiving respectively. The energy for transmission $E_t$ is the energy required to transmit each packet and computed as:

$$E_t = P_t + T_t,$$ (2)

where, $P_t$ is the number of packets transmitted and $T_t$ is the time required to transmit each packet. The Receiving Energy Re is the energy required to receive each packet. This can be calculated as:

$$E_r = P_r + T_t,$$ (3)

where, $P_r$ is the number of packets received, $T_r$ is the time required to receive each packet.

## 5.3. ENERGY DENSITY

The Energy density, $E_d$ of a given node can be calculated as the ratio of sum of Residual energy of all the nodes within the radio coverage disc to the area of the radio coverage disc of the entire network. The Energy density, $E_d$ is calculated as follows:

$$E_d(t) = \frac{R_e(i) + R_e(i+1) + R_e(i+2) + \cdots + R_e(n)}{Distance(i,k) * Coverage area},$$ (4)

where $E_d(t)$ Energy density of node $i$, $Distance(i,k)$ is the distance from node $i$ to local cluster head, where $k = CH_i$. The $Coverage area$ is the range of the area around a node that can transmit to its neighbour nodes. This Coverage Area is in terms of meters.

## 5.4. CLUSTER GENERATOR

Clusters are generated in the second phase that accepts randomly deployed sensor nodes. The phase generates number of clusters based on the maximum number of sensor nodes. Let, $n_1, n_2, n_3, \ldots n_m$, are the random nodes, where $n_m$ is a maximum number of nodes deployed in the network. The cluster generation can be done by grouping different nodes and each cluster can be represented as,

$$C_i = n_1, n_2, n_3, \ldots n_n,$$

where $C_i$ is the cluster identifier, $i = 1,2,3,\ldots n$ and $n_n$ is the maximum number of nodes in each cluster.

## 5.5. SELECTION OF CLUSTER HEAD

The third phase involves the selection of cluster head that is carried out in two steps. The first step involves the local cluster head selection and second step involves in the selection of the remaining cluster heads. The local cluster head, $CH_l$ will be selected randomly from any one of the clusters among $C_i$. The $CH_l$ is selected based on two parameters $R_e(i)$ and $Cost(i)$. Let,

$$CH_l = max\big(R_e(i)\big) \,\&\&\, Cost(i),$$

where $R_e(i)$ and $Cost(i)$ are the residual energy and cost of the nodes. The residual energy $R_e(i)$ can be computed using equation (1) and $Cost(i)$ is generated for each node through random function. In the next step, the remaining cluster heads are selected based on the Energy density of all the nodes except the nodes present in the local cluster, $C_i$. The Energy density of each node can be calculated as $E_d(i)$ using equation (4), where $E_d(i)$ is the energy density of each node i. The node which has highest Energy density, among other nodes within the cluster becomes the cluster head. The cluster head selection process is represented as follows.

$$CH_i = max\big(E_d(i)\big),$$

This is repeated until all the clusters select their cluster heads and for every event generation at each node. The two step cluster head selection technique is presented in Table 1.

Table 1. Algorithm for Cluster Head Selection

## 5.6. ON-DEMAND SHORTEST PATH (ODSP)

The fourth phase routes the packet from source node to the sink *via* the shortest path. It is required to calculate the depth from source to sink through different clusters.

$$C_i = d(i),$$

where $C_i$ is the i$^{th}$ cluster and $d(i)$ is the depth of cluster $C_i$. Events generated in any one of the node is broadcast by the Route request message to the sink through its neighbour cluster heads. In turn, the Sink replies through the shortest path to the source node based on the Depth of the cluster head $d(i)$. The process of finding the shortest path dynamically is called as On-Demand Shortest Path (ODSP) and represented by,

$$ODSP = \min\big(d(i)\big).$$

Finally, the packet is transmitted from the source node to the sink in the fifth phase. EDOCR is presented in Table 2. The clusters are formed and cluster head is selected as described earlier. The route between source and sink is established dynamically on-demand. Initially, the source and sink exchange messages to establish the shortest path with minimum depth $d$. After successful negotiation, packets are transmitted from source to the sink. *Send_Route_Request_Message()* and *Send_Route_Reply_Message()* are the two functions that performs the broadcasting of route request message from the source node to the sink and reply message from the sink to the source node respectively.

Table 2. EDOCR Algorithm

**Algorithm 2** EDOCR Algorithm
**input** : number of nodes - N , number of clusters -M ,
nodes
    **Step 1 :** Cluster Head Selection
        *Cluster_Head_Selection()*
    **Step 2 :** Send Route Request Message
        *Send_Route_Request_Message()*
    **Step 3 :** Calculate Shortest Depth
        **for** *i = 0* to *M* **do**
          **for** *j = 0* to *M* **do**
            **if** $d(CH_i) < d(CH_j)$ **then**
              select $CH_i$
            **else**
              select $CH_j$
            **end if**
          **end for**
        **end for**
    **Step 4 :** Send Route Reply Message
        *Send_Route_Reply_Message()*
    **Step 5 :** Forward packets through the shortest path with minimum depth d return by sink
    **Step 6 :** Repeat above step until all the packets move from source to sink

## 6. SIMULATION AND PERFORMANCE ANALYSIS

The EDOCR protocol is simulated using NS-2 Simulator. Simulation results in terms of Network Lifetime, Packet Delivery Ratio, Packet Drop and Network Throughput are depicted in the following section. Our simulation set up includes 50 sensor nodes which are randomly deployed and distributed over the area of 1300 * 1000 meters as shown in Fig. 2. The clusters for these nodes are depicted in Fig. 3.

## 6.1. PERFORMANCE METRICS

The EDOCR algorithm is analysed with respect to the following performance metrics.

*Network Lifetime (NL):* It is defined as the maximum total time duration for which the network provides fair connectivity among all the nodes without network partition.

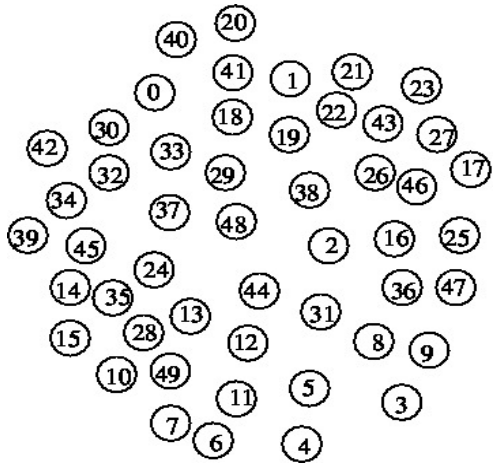

Figure 2. Random Deployment of Sensor nodes.

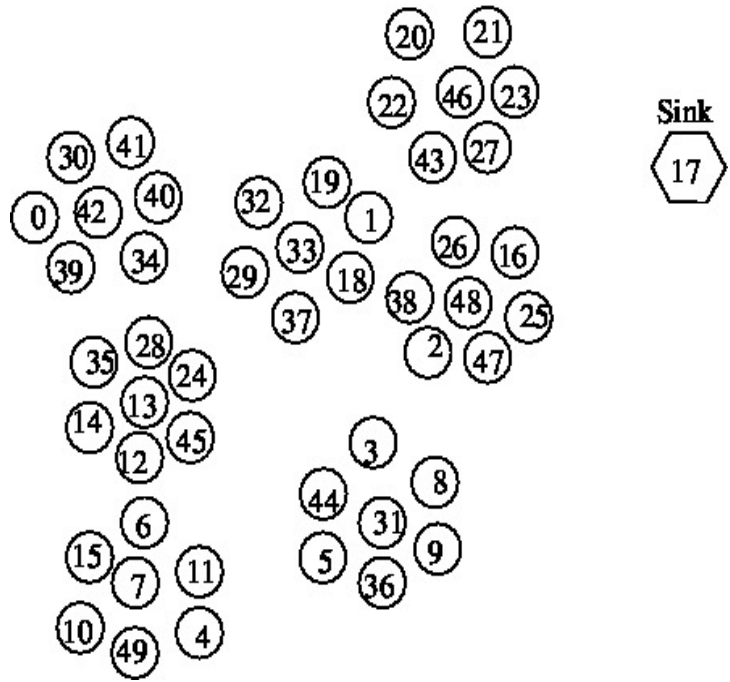

Figure 3. Cluster Formation

***Packet Delivery Ratio (PDR)*:** It is the ratio of actual packet received by the sink to the total number of packets sent from the source.
***Packet Drop (PD)*:** The number of packets lost during the packet transmission over a stipulated time period.
***Network Throughput (NT)*:** It is defined as the rate of packet transmission in the network per unit of time.

## 6.2. PERFORMANCE ANALYSIS

Network size, Coverage area, Clusters node distribution, Packet size, Initial energy and Simulation Time values are given in Table 3.

Table 3. Simulation Parameter

| Parameter | Values |
|---|---|
| Network Size | 1300m*1000m |
| Number of nodes | 50 |
| Number of clusters | 7 |
| Node Distribution | Random |
| Initial Energy | 1J |
| Data Packet Size | 64 |
| Location of Sink Node | 1004.5m*619.613m |
| Coverage area for an individual Sensor node | 3m |
| Simulation time | 2000s |

Table 4 shows the performance metrics Network Lifetime, Packet Delivery Ratio, Packet Drop and Network Throughput for the the proposed protocol EDOCR in comparison with TSCHS and EBRP.

Fig. 4. depicts the comparative value of Network Lifetime with our protocol (EDOCR) with the earlier protocols two Step Cluster Head Selection (TSCHS) and Energy Balanced Routing Protocol (EBRP). All the three protocol exhibit the same level of energy utilization till 6000ms. It is observed that between the time period 6000ms to 14000ms, the energy utilization of the three protocols starts differing. The depletion of energy is much slower in EDOCR protocol than the other two protocols. This is an account of cumulative impact of cluster head selection on the basis of energy density and on-demand shortest path routing resulting in uniform utilization of energy among all the nodes of the network that enhances the lifetime of the network. It is observed that the Network Lifetime increases by 18% over the other two protocols(Table 4).

Table 4. Comparison values of NL and PDR

| Simulation Time | Network Lifetime | | | Packet Delivery Ratio | | |
|---|---|---|---|---|---|---|
| | EDOCR | TSCHS | EBRP | EDOCR | TSCHS | EBRP |
| 2000 | 0.98 | 0.98 | 0.98 | 0.82 | 0.73 | 0.78 |
| 4000 | 0.98 | 0.98 | 0.98 | 0.83 | 0.75 | 0.80 |
| 6000 | 0.98 | 0.98 | 0.98 | 0.86 | 0.78 | 0.81 |
| 8000 | 0.97 | 0.94 | 0.96 | 0.88 | 0.80 | 0.84 |
| 10000 | 0.90 | 0.92 | 0.88 | 0.92 | 0.84 | 0.86 |

| 12000 | 0.82 | 0.86 | 0.79 | 0.94 | 0.86 | 0.88 |
| 14000 | 0.76 | 0.75 | 0.72 | 1 | 0.88 | 0.90 |

Packet Delivery Ratio is depicted in Fig. 5. and Table 4. It is observed that the packet delivery ratio is lower in TSCHS and EBRP than the proposed protocol EDOCR. While the packet drop ratio is almost same in the earlier protocols, our protocol has higher packet delivery on account of low node failure and high residual energy at any instant of time. The performance of packet delivery ratio further increases with simulation time. It is seen that there is an increase of 25% in packet deliver ratio in comparison to other two protocols.

Table 5. Comparison values of PD and NT

| Simulation Time | Packet Drop | | | Network Throughput | | |
|---|---|---|---|---|---|---|
| | EDOCR | TSCHS | EBRP | EDOCR | TSCHS | EBRP |
| 2000 | 0.13 | 0.18 | 0.15 | 4.3 | 4.0 | 4.2 |
| 4000 | 0.15 | 0.20 | 0.17 | 4.5 | 4.1 | 4.3 |
| 6000 | 0.2 | 0.25 | 0.3 | 4.6 | 4.2 | 4.4 |
| 8000 | 0.25 | 0.33 | 0.2 | 4.7 | 4.5 | 4.4 |
| 10000 | 0.3 | 0.42 | 0.35 | 4.75 | 4.52 | 4.45 |
| 12000 | 0.32 | 0.44 | 0.38 | 4.8 | 4.56 | 4.5 |
| 14000 | 0.34 | 0.46 | 0.43 | 4.9 | 4.67 | 4.56 |

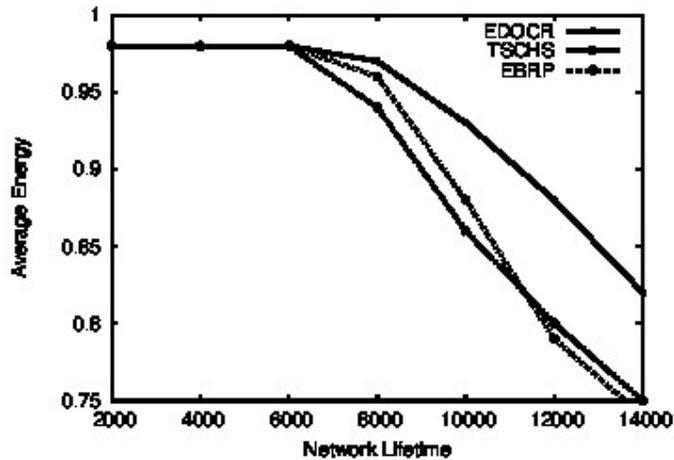

Figure 4. Network Lifetime

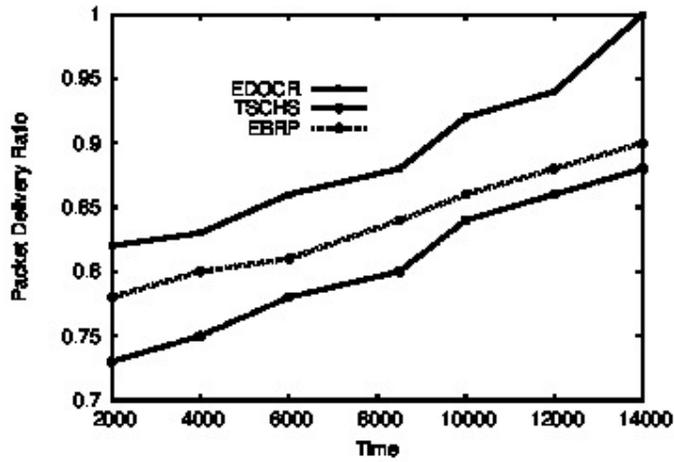

Figure 5. Packet Delivery Ratio

Fig. 6. illustrates the number of packet dropped in EDOCR, TSCHS and EBRP. A packet is dropped on account of congestion or failure of node to transmit the packet due to low residual energy. Since both these parameters are taken care of through careful selection of cluster heads, the packet drop ratio is lower in EDOCR than other two protocols. The packet drop ratio difference in three protocols is low in the beginning. Later, the rate of packet drops increase abruptly in TSCHS and linearly in EBRP after 4000ms. Our protocol exhibits the rate of packet drops linearly with comparatively lower percentage (22%).

The throughput of the network in EDOCR is shown in Fig. 7. Uniform energy utilization coupled with high packet delivery ratio and low packet drop in EDOCR has resulted in higher throughput than TSCHS and EBRP. There is an increase in throughput of 38% in our protocol and the throughput profile is shown in Table 5.

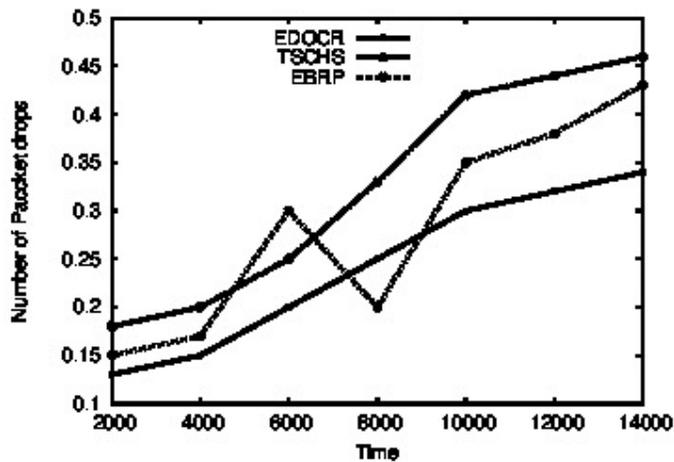

Figure 6. Packet Drop

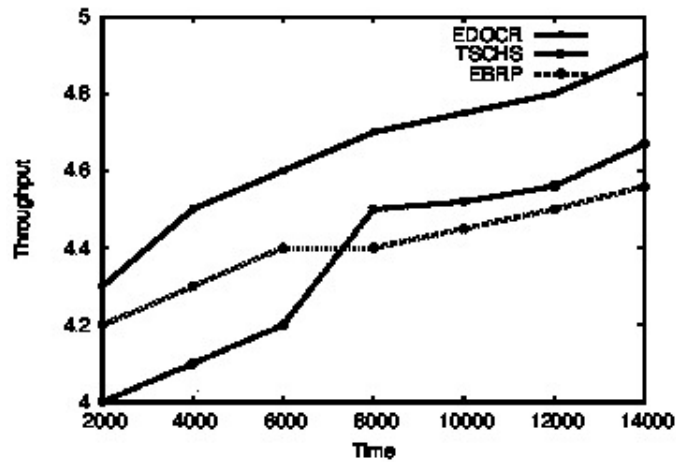

Figure 7. Network Throughput

# 7. CONCLUSIONS

Balanced Energy utilization is one of the important parameters in increasing the lifetime of the WSNs. Cluster head selection and on-demand routing are two critical issues. The proposed algorithm EDOCR shows that the cluster head selection based on the energy density and the residual energy is more efficient and effective than other parameters. We have followed a unique energy density calculation, based on the parameters such as average residual energy of neighbour nodes and itself, distance from local cluster head and coverage area of each node. Such a method of cluster head selection supports balanced energy utilization and increase in throughout of the network. The shortest path calculation is based on On-Demand approach which considers depth parameter from source to sink. The cumulative efficiency of the Two steps cluster head selection coupled with shortest path On-Demand routing has increased the network lifetime. It is clearly observed that our algorithm performs better than earlier algorithms with respect to uniform utilization of energy, lifetime, packet delivery ratio and throughput. This work can be extended to mobile sinks to reduce latency and further increase the lifetime of the network.

**Authors**

**B M Thippeswamy** is an Assistant Professor and Head in the Department of Computer Science and Engineering at Sambhram Institute of Technology, Bangalore, India. He obtained his B.E in Computer Science and Engineering from Mysore University and M.E degrees in Computer Science and Engineering from Bangalore University, Bangalore. He is presently pursuing his Ph.D programme in the area of Wireless Sensor Networks in JNTU Ananthapur, India. His research interest is in the area of Wireless Sensor Networks. 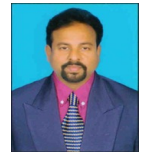

**Reshma S** is a Lecturer in the Department of Computer Science and Engineering at Sambhram Institute of Technology, Bangalore, India. She received her Bachelors degree in Computer Science and Engineering from Visvesvaraya Technological University and Master of Technology from Visvesvaraya Technological University, Regional Center, Bangalore. Her research interest is in the area of Wireless Sensor Networks. 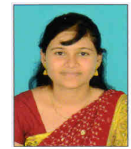

**Shaila K** is an Professor and Head in the Department of Electronics and Communication Engineering at Vivekananda Institute of Technology, Bangalore, India. She obtained her B.E in Electronics and M.E degrees in Electronics and Communication Engineering, and Ph.D degree from Bangalore University, Bangalore. Her research interest is in the area of Sensor Networks, Adhoc Networks and Image Processing. 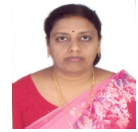

**Venugopal K R** is currently the Principal, University Visvesvaraya College of Engineering, Bangalore University, Bangalore. He obtained his Bachelor of Engineering from University Visvesvaraya College of Engineering. He received his Masters degree in Computer Science and Automation from Indian Institute of Science Bangalore. He was awarded Ph.D. in Economics from Bangalore University and Ph.D. in Computer Science 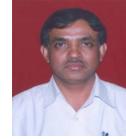 from Indian Institute of Technology, Madras. He has a distinguished academic career and has degrees in Electronics, Economics, Law, Business Finance, Public Relations, Communications, Industrial Relations, Computer Science and Journalism. He was a postdoctoral research scholar at University of Southern California, USA. He has authored and edited 35 books on Computer Science and Economics, which include Petrodollar and the World Economy, C Aptitude, Mastering C, Microprocessor Programming, Mastering C++ and Digital Circuits and Systems etc.. During his three and half decades of service at University Visveraya College of Engineerng. He has over 300 research papers to his credit. His research interests include Computer Networks, Wireless Sensor Networks, Parallel and Distributed Systems, Digital Signal Processing and Data Mining.

**S S Iyengar** S S Iyengar is currently Ryder Professor, Florida International University, USA. He was Roy Paul Daniels Professor and Chairman of the Computer Science Department at Louisiana State University. He heads the Wireless Sensor Networks Laboratory and the Robotics Research Laboratory at LSU. He has been involved with research in High Performance Algorithms, Data Structures, Sensor Fusion and Intelligent 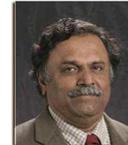 Systems, since receiving his Ph.D degree in 1974 from MSU, USA. He is Fellow of IEEE and ACM. He has directed over 40 Ph.D students and 100 Post Graduate students, many of whom are faculty at Major Universities worldwide or Scientists or Engineers at National Labs/Industries around the world. He has published more than 500 research papers and has authored/co-authored 6 books and edited 7 books. His books are published by John Wiley & Sons, CRC Press, Prentice Hall, Springer Verlag, IEEE Computer Society Press etc.. One of his books titled Introduction to Parallel Algorithms has been translated to Chinese.

**L M Patnaik** is currently Honorary Professor, Indian Institute of Science, Bangalore, India. He was a Vice Chancellor, Defense Institute of Advanced Technology, Pune, India and was a Professor since 1986 with the Department of Computer Science and Automation, Indian Institute of Science, Bangalore. During the past 35 years of his service at the Institute he has 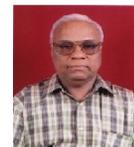 over 700 research publications in refereed International Journals and refereed International Conference Proceedings. He is a Fellow of all the four leading Science and Engineering Academies in India; Fellow of the IEEE and the Academy of Science for the Developing World. He has received twenty national and international awards; notable among them is the IEEE Technical Achievement Award for his significant contributions to High Performance Computing and Soft Computing. His areas of research interest have been Parallel and Distributed Computing, Mobile Computing, CAD, Soft Computing and Computational Neuroscience.